\documentclass[mathleft
]{an}
\usepackage{graphicx}
\usepackage{times}
\usepackage{color}
\usepackage{revsymb}        
\usepackage{amsmath}        
\usepackage{txfonts}
\usepackage{natbib}
\bibpunct{(}{)}{,}{a}{}{,}
\newcommand{\aap}{Astron. Astrophys.}          
\newcommand{\prd}{Phys. Rev.~D}          

\begin{document}

\Pagespan{1}{}
\Yearpublication{2010}%
\Yearsubmission{2010}%
\Month{11}%
\Volume{999}%
\Issue{88}%


\title{2D non-perturbative modeling of oscillations \\
        in rapidly rotating stars}

\author{R-M. Ouazzani\inst{1}\fnmsep\thanks{Corresponding author:
  \email{rhita-maria.ouazzani@obspm.fr}\newline}, M-A. Dupret\inst{2}, M.J. Goupil\inst{1},D.R. Reese\inst{1}
}

\titlerunning{2D Non Perturbative Modeling of Oscillations}
\authorrunning{R-M. Ouazzani}
\institute{LESIA, UMR8109, Universit\'e Pierre et Marie Curie, Universit\'e Denis Diderot, Observatoire de Paris, 92195 Meudon Cedex, France
\and 
Institut d'Astrophysique et de Geophysique, Fac. Sc., Universit\'e de Li\`ege}

\received{30 May 2005}
\accepted{11 Nov 2005}
\publonline{later}


\abstract{%
 We   present and discuss  results of a  recently developped  two dimensional
non-perturbative method to compute accurate adiabatic oscillation modes of
rapidly rotating stars .  The 2D calculations fully take into account the
centrifugal distorsion of the star while the non-perturbative method includes
the full influence of the Coriolis acceleration. These characteristics allows us
to compute oscillation modes of rapid rotators - from high order
p-modes in $\delta$Scuti stars, to low order p- and g-modes in $\beta$ Cephei or Be
stars. 
}

\maketitle

\section{Introduction}

It is a well known fact that rotation plays a key role in stellar evolution.
From early stages of stellar formation to the final steps of evolution,
rotation generates various dynamical processes such as meridional circulation
and rotationnally induced turbulence which drive chemical element mixing and
transport of angular momentum \citep[for example][]{Maeder2009}. These
processes are not fully understood and are still poorly modeled, but
asteroseismology can provide important constraints provided the
effects of rotation on stellar pulsation are better
understood.  In rotating stars, the centrifugal
acceleration breaks the spherical symmetry -- causing
distorsion -- and the resonant cavity of the modes is modified. The Coriolis
acceleration enters the equation of motion and affects the motion of the waves
and the frequencies of the normal modes.  For slow rotators, the effects
of rotation on oscillation frequencies have been extensively investigated with
perturbative methods \citep[see][and references therein]{Saio81,GT90,DG92,S98}.
In this approach, the angular rotation velocity
$\Omega$ is considered as small compared to the oscillation
frequencies, thereby allowing their expansion as a power
series in $\Omega$. Perturbation methods cease to be valid whenever the
rotation frequency is no longer negligible in front of the break-up frequency
($\sqrt{GM/R^3}$) or the oscillation frequency. Then for moderate to rapid
rotators a non-perturbative treatment is necessary. In the non-perturbative
approach, the pulsation equations are projected onto the spherical
harmonic basis. The effects of the Coriolis acceleration
and the stellar distorsion cause a coupling between the different spectral
components, and the eigenvalue problem must be solved directely by a
two-dimensionnal method. Such an approach has been applied to g and r-modes for uniformally rotating stars under the Cowling approximation \citep{Lee1987}, for acoustic
modes in uniformally rotating polytropes \citep{Reese2006,Lignieres2006}, in
uniformally rotating ZAMS models \citep{Lovekin2008}, and in
differentially rotating ZAMS models with a conservative rotation law
\citep{Lovekin2009, Reese2009}.  We present here a 2D non-perturbative code
which allows us to calculate adiabatic oscillations for
the whole frequency range from high order g-modes to high order p-modes.
Particular care has been taken so as to be able to compute
pulsations for all types of stellar models. In the oscillation code, no
hypotheses have been made on the fluid microphysics -- non polytropic,
non barotropic -- the rotation profil is free -- not necessarily
conservative -- it can be differential in radius and in latitude. The paper is
organised as follows: in the next section, the formalism is explained. In
section 3, we describe the numerical method used to solve the eigenfunction
problem. A conclusion and perspectives follow.

\section{The formalism}
For the computation of pulsations, the stellar structure is reduced to its
dynamical behavior. We compute oscillation modes as the adiabatic
response of the structure to small perturbations -- i.e. of the density,
pressure, gravitationnal potential and velocity field -- using the eulerian
formalism.  The velocity field of the equilibrium structure is only due to
solenoidal rotation:
\begin{align}
\vec{\rm v_0}&=\vec{\Omega} \times \vec{\rm r}\\
\hbox{where} \hspace{0.5cm} \vec{\Omega}=\Omega(\rm r,\theta) \cos(\theta)& \, \vec{\rm e_{r}} -  \,\Omega(\rm r,\theta) \sin(\theta)  \, \vec{\rm e_{\theta}}
\end{align}
$\vec{\rm e_{r}}$ and $\vec{\rm e_{\theta}}$ being the classical spherical basis
vectors. Then the equations describing the oscillations of a self rotating
fluid are the perturbed equations of motion:
\begin{eqnarray}
\label{motion}
\rho_{0}  \left( \frac{\partial\mathbf{\rm v'}}{\partial \rm t} + (\mathbf{\rm v_{0}}.\mathbf{\nabla})\mathbf{\rm v'} + (\mathbf{\rm v'}.\mathbf{\nabla}) \mathbf{\rm v_{0}}\right) &+& \rho' (\mathbf{\rm v_{0}}.\mathbf{\nabla})\mathbf{\rm v_{0}}  \nonumber \\
=-\mathbf{\nabla}\rm p'-\rho'\mathbf{\nabla}\Phi_{0}-\rho_{0}\mathbf{\nabla}\Phi'
\end{eqnarray}
The linearised continuity equation:
\begin{equation}
\label{continuity}
\left(\frac{\partial}{\partial \rm t} + \Omega \frac{\partial}{\partial \phi}\right) \rho'+\mathbf{\nabla}.\left(\rho_{0}\mathbf{\rm v'}\right)=0
\end{equation}
The Poisson equation for the perturbed gravitational potential:
\begin{equation}
\label{poisson}
\nabla^2\,\Phi' = \, 4 \pi \rm G \rho'
\end{equation}
and the perturbed adiabatic relation:
\begin{equation}
\label{energy}
\left( \frac{\partial}{\partial \rm t}+\Omega \frac{\partial}{\partial \phi} \right) \left( \frac{\rho'}{\rho_0}-\frac{\rm p'}{\Gamma_1 \rm p_0}\right) + \mathbf{\rm v'}. \left( \nabla \ln \rho_0 - \frac{1}{\Gamma_1} \nabla \ln \rm p_0 \right)=0
\end{equation}
By adding an auxiliary equation for the derivative of the
gravitationnal potential $d\Phi'=\partial \Phi'/\partial \zeta$, the problem is
reduced to a set of first order differential and algebraic equations. Together
with specific boundary conditions, we get an eigenvalue problem,
the eigenvalues of which are the oscillation
frequencies, and the eigenfunctions of which are the
eulerian perturbations of density, pressure, gravitational
potential, its radial derivative, and the three spatial components of
the perturbed velocity field.

\subsection{Spheroidal geometry}
Due to the distorted shape caused by rotation, we chose to use a spheroidal
coordinate system. This system \citep[found by][]{Bonazzola1998} is convenient
for setting up proper boundary conditions. As done in \cite{Reese2006}, $\zeta$ is
defined as the radial coordinate, and is related to the spherical r coordinate
by:
\begin{itemize}
\item In the stellar interior $\zeta \, \in \left[ 0;1\right]$, domain V:
\begin{align}
\rm r(\zeta,\theta) = (1-\epsilon) \zeta + \frac{5\zeta^3 - 3\zeta^5}{2} \left( \rm R_s(\theta) - 1 + \epsilon \right)
\end{align}
\item In the outer domain $\rm V_2$,  $\zeta \, \in \left[ 1;2\right]$:
\begin{align}
\rm r(\zeta,\theta) = \, & \, 2 \, \epsilon \, + \, (1-\epsilon) \zeta \nonumber \\
&+ \left( 2\zeta^3 - 9\zeta^2+12 \zeta -4 \right)  \left( \rm R_s(\theta) - 1 + \epsilon \right)
\end{align}
\end{itemize}

\noindent
where $\theta$ is the colatitude, $\epsilon = 1\, - \, R_{pol}/R_{eq}$
the flatness and $\rm R_s(\theta)$ the radius at the surface.
With this mapping, the surface of the star is given by $\zeta=1$, which
is very convenient for avoiding discontinuities at the
stellar surface. At the center, surfaces of constant $\zeta$ tend
to be spherical, so that central regularity conditions are simplified. In the outer
region, iso-$\zeta$ surfaces regain a spherical shape at $\zeta=2$.
\begin{figure}[h!]
 \centering{
\includegraphics[scale=0.3]{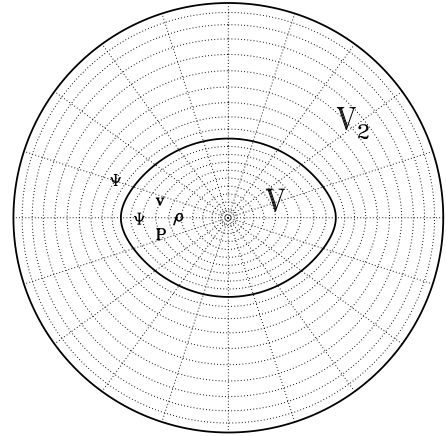}
\caption{Coordinate system used in computing the pulsation modes. The domain
$\rm V$ corresponds to the star itself. $\rm V_2$ encompasses the star, its
outer limit being a sphere of radius $\rm r = 2$ (twice the equatorial radius).}}
\end{figure}

\subsection{Boundary conditions}

In order to complete the eigenvalue problem defined by the four equations
Eq.(\ref{motion}), Eq.(\ref{continuity}), Eq.(\ref{poisson}), Eq.(\ref{energy}),
it is necessary to specify a number of boundary conditions. 

\noindent
At the center of the star, where the mapping is almost spherical, the
requirement is that the velocity field components are regular.
This condition is easily expressed for functions expanded over the
spherical harmonic basis. Concerning the scalar quantities, we ensure
that the radial component associated with the spherical harmonic $\rm Y_{\ell}^{\rm m}$
behaves like $\rm r^{\ell}$ (see Sect.3.2). 

\noindent
At the surface of the star, the boundary condition is satisfied by
means of the stressless condition $\delta \rm p'=0$, which is
easily expressed on the iso-$\zeta$ surface $\zeta=1$. 

\noindent
It is also necessary to impose a condition on the gravitational potential so
as to ensure that it goes to zero at infinity.  With the mapping
described in the former section, we can safely impose the classical
condition on the different harmonic components of $\Phi'$ on the outer border
of $\rm V_2$, at $\zeta=2$.

\subsection{Change of variable}
In order to have solutions with a good behavior at the surface of the star, we choose to use $\pi'=\rm p'/\rho_0$ rather
than the pressure perturbation directly. 

\section{Numerical method}
In order to isolate the radial, spheroidal and toroidal components of
the fluid's motion, we take the radial componant of the equation
of motion (Eq.\ref{motion}), its divergence and the radial part of its curl.  We then get a set of $7$ equations and $7$ unknowns which
are: the perturbed velocity vector, ($\pi'$), the density
perturbation ($\rho'$), the perturbation of the gravitationnal
potential ($\Phi'$) and its derivative ($\rm d \Phi'/\rm d\zeta$), along with
the related boundary conditions. 

\subsection{Projection onto the spherical harmonics}
To solve this eigenvalue problem, we develop all the variables onto a serie of
spherical harmonics, following  \cite{Rieutord1987}:\\
For the velocity field perturbation:
\begin{align}
\label{spect_vitesse}
\rm v'_{\zeta}(\zeta,\theta,\phi)&=i \sum_{\ell_2 \geq \mid \rm m \mid}^{+ \infty} \rm u_{\ell_2}(\zeta)  \rm Y_{\ell_2}^{\rm m}(\theta,\phi) \\
\rm v'_{\theta}(\zeta,\theta,\phi)&=i \sum_{\ell_2 \geq \mid \rm m \mid}^{+ \infty} \left( \rm v'_{\ell_2}(\zeta)  \frac{\partial \rm Y_{\ell_2}^{\rm m}(\theta,\phi) }{\partial \theta} + \rm w'_{\ell_{2p}}(\zeta)  \frac{\rm m}{\sin \theta} \rm Y_{\ell_{2p}}^{\rm m}(\theta,\phi) \right) \nonumber \\
\rm v'_{\phi}(\zeta,\theta,\phi)&= - \sum_{\ell_2 \geq \mid \rm m \mid}^{+ \infty} \left( \rm v'_{\ell_2}(\zeta)  \frac{\rm m}{\sin \theta} \rm Y_{\ell_{2}}^{\rm m}(\theta,\phi) +  \rm w'_{\ell_{2p}}(\zeta)  \frac{\partial \rm Y_{\ell_{2p}}^{\rm m}(\theta,\phi) }{\partial \theta} \right) \nonumber
\end{align}
where $\ell_{2p}=\ell_2+(-1)^{p\rm }$. For any scalar variable ($\pi'$, $\rho'$, $\Phi'$, $d\Phi'$):
\begin{equation} 
\label{spect_scalaire}
f'(\zeta,\theta,\phi)=\sum_{\ell_2 \geq \mid \rm m \mid}^{+ \infty} \rm f'_{\ell_2}(\zeta) . \rm Y_{\ell_2}^{\rm m}(\theta,\phi)
\end{equation}
where $\rm Y_{\ell}^{\rm m}$ is the spherical harmonic of degree $\ell$
and azimuthal order m, and $\rm f'_{\ell}(\zeta)$ the radial
functions that are to be determined. We include these spectral developments
Eq.(\ref{spect_vitesse}) and Eq.(\ref{spect_scalaire}) into the equations system
Eq.(\ref{motion}) to Eq.(\ref{energy}). We get a linear partial differential
equations system in terms of the variables $\rm u'_{\ell_2}$, $\rm v'_{\ell_2}$,
$\rm w'_{\ell_2}$, $\pi'_{\ell_2}$, $d\Phi'_{\ell_2}$, $\Phi'_{\ell_2}$ and
$\rho'_{\ell_2}$ which can formally be written as:
\begin{equation}
\label{Eql2}
  \sum_{\ell_2 \geq \mid \rm m \mid} \, \rm E\left( \rm Y_{\ell_2}^{\rm m}(\theta,\phi) \right) \, = \, 0
\end{equation}
In the general case -- where rotation breaks the spherical symmetry -- the
problem is not separable in $\zeta$ and $\theta$. As a result, a coupled equations system
is obtained by projecting Eq.(\ref{Eql2}) onto the spherical harmonics basis:
\begin{equation}
\label{principe_decomp}
\forall \, \ell_1 \, \geq \, | \rm m | \, , \,\sum_{\ell_{2} \geq \mid \rm m \mid}^{+ \infty} \int \frac{\sin \theta \, d\theta \, d\phi}{4 \pi} \, \rm E(\rm Y_{\ell_{2}}^{\rm m}) \, Y_{\ell_{1}}^{\rm m \, *} \, = \, 0
\end{equation}
where $\rm Y_{\ell_{1}}^{\rm m \, *}$ is the complex conjugate
of $\rm Y_{\ell_{1}}^{\rm m}$. Finally, to get a finite and equal number of
equations and variables, we truncate the series at the 2M$^{th}$ term. Given
that the equilibrium model is symmetrical with respect to the equator, half of
the terms are dropped in the projection, and the system only couples terms with
the same symmetry, i.e. parity. The selection rules operating here are:
\begin{align}
\ell &= | \rm m| + 2(\rm j-1)+\rm p, \hspace{0.4cm}  \rm j \in \left[ 1:\rm M \right]  
\end{align}
\begin{itemize}
\item[ ] $\rm p=0$ if m and $\ell$ are of the same parity,
\item[ ]$\rm p=1$ otherwise.
\end{itemize}
\noindent
We then are able to solve the eigenvalue problem separately for the even
and odd eigenfunctions, and for a given azimuthal order.

\subsection{Radial behavior}
In order to ensure a proper regular behavior at the center of the star, and to
avoid numerical convergence problems, we scale the radial components of
the spectral decomposition by the appropriate powers of $\zeta$.
If we suppose a priori the regularity of the scalar variables $\rm f_{\ell}$
($\pi_{\ell}^{' }$, $\rho_{\ell}^{'}$ and $\Phi_{\ell}^{'}$), following
Nikiforov and Uvarov (1983), these radial functions satisfy:
\begin{equation}
\rm f_{\ell}(\zeta) \sim \zeta^{\ell} \hspace{0.4cm} \hbox{when} \hspace{0.4cm} \zeta \rightarrow 0
\end{equation}
For the velocity field, the same treatment gives:
\begin{equation}
\rm u_{\ell}^{'}\, \sim \, \zeta^{\ell-1}, \hspace{0.4cm} \rm v_{\ell}^{'}\, \sim \, \zeta^{\ell-1} \hspace{0.4cm} \hbox{and} \hspace{0.4cm} \rm w_{\ell}^{'}\, \sim \, \zeta^{\ell}
\end{equation}
This leads to the following scaling:
\begin{align}
\pi'_{\ell} \, = \zeta^{\ell} \, \tilde{\pi_{\ell}}'  \hspace{0.5cm} \Phi'_{\ell} \, = \, \zeta^{\ell} \, \tilde{\Phi_{\ell}}' & \hspace{0.5cm} \rho'_{\ell} \, = \, \zeta^{\ell} \, \tilde{\rho_{\ell}}' \hspace{0.5cm} \rm d\Phi'_{\ell} \, = \, \zeta^{\ell-1} \tilde{\rm d\Phi_{\ell}}' \nonumber \\
\rm u'_{\ell} \, = \, \zeta^{\ell-1} \, \tilde{\rm u_{\ell}}' \hspace{0.5cm} \rm v'_{\ell} &= \zeta^{\ell-1} \, \tilde{\rm v_{\ell}}' \hspace{0.5cm} \rm w'_{\ell} \, = \, \zeta^{\ell} \, \tilde{\rm w_{\ell}}'
\end{align}

\subsection{Final eigenvalue system}

In the set of equations, 4 are differential equations for the
variables $u'$, $\pi'$, $\Phi'$ and $d\Phi'$, with respect to the radial
coordinate $\zeta$, and 3 are not:
\begin{eqnarray}
\frac{\rm dy_1}{\rm d \zeta} &=& (\rm A_{11}+ \delta \sigma \rm A_{12}) \rm y_1 + (\rm A_{21}+\delta \sigma \rm A_{22}) \rm y_2 \\
0&=&(\rm B_{11}+\delta \sigma \rm B_{12}) \rm y_1 + (\rm B_{21}+ \delta \sigma \rm B_{22}) \rm y_2
\label{substit}
\end{eqnarray}
\noindent
where $\sigma = \sigma_0 + \delta \sigma$. $y_1$ and $y_2$ are the column
vectors -- with 4 M and 3 M components respectively -- containing the
unknown coefficients of the spectral decomposition:
\begin{eqnarray}
\rm y_1&=&(\tilde{\pi}'_{\ell_1}, ..., \tilde{\pi}'_{\ell_M},\tilde{\rm d\Phi}'_{\ell_1}, ..., \tilde{\rm d\Phi}'_{\ell_M}, \tilde{\Phi}'_{\ell_1}, ..., \tilde{\Phi}'_{\ell_M}, \tilde{\rm u}'_{\ell_1}, ..., \tilde{\rm u}'_{\ell_M}) \nonumber \\
\rm y_2&=& (\tilde{\rm v}'_{\ell_1}, ... \tilde{\rm v}'_{\ell_M}, \tilde{\rm w}'_{\ell_1}, ... \tilde{\rm w}'_{\ell_M}, \tilde{\rho}'_{\ell_1}, ... \tilde{\rho}'_{\ell_M}) \nonumber \\
\hbox{where} \nonumber \\
\ell_1 \, &=& \, |\rm m|+\rm p \, \hspace{0.2cm} \hbox{and} \hspace{0.2cm}\,  \ell_M \, = \, |\rm m|+2(\rm M-1)+\rm p 
\end{eqnarray}

Using (\ref{substit}), a matrix inversion allows to express $y_2$ as a linear
function of $y_1$. The system is then reduced to a differential system of 4
independent variables:
\begin{equation}
\Rightarrow \, \frac{\rm dy_1}{\rm d\zeta} \simeq \left( \rm A+ \delta \sigma. \rm A_{\delta \sigma} \right) \rm y_1 
\end{equation}

\subsection{Radial resolution: finite differences}
Considering two consecutive layers i and i+1, such that \\ $\zeta(\rm i+1)-\zeta(\rm i)=\rm h$, a Taylor development of any function y gives \citep{Scuflaire2008}:
\begin{align}
\label{diff_scuf}
&\rm y(i)\, + \, \frac{\rm h}{2} \, \frac{\rm dy}{\rm d\zeta}(i) \, + \, \frac{\rm h^2}{12} \, \frac{\rm d^2y}{\rm d\zeta^2}(i)\, = \nonumber \\
& \rm y(i+1) \, - \, \frac{\rm h}{2} \,  \frac{\rm dy}{\rm d\zeta}(i+1) \, + \, \frac{\rm h^2}{12} \,  \frac{\rm d^2y}{\rm d\zeta^2}(i+1)  \, +  \, o\left(\rm h^5\right)
\end{align}
We can develop a $5^{\rm th}$ order finite differences scheme that only involves
two consecutive layers ($\zeta_i$ and $\zeta_{i+1}$).\\
We then obtain a global eigenvalue system:  
\begin{equation}
\label{eigenpb}
\rm AA \, \rm Y = \delta \sigma  ~\rm AA_{\delta \sigma} \, \rm Y
\end{equation}
Where AA and $\rm AA_{\delta \sigma}$ are block diagonal matrices. The block i
couples the layers i and i+1 and:
\begin{equation}
\rm Y = \begin{pmatrix} \rm y_1(1) \\ \vdots \\ \rm y_1(N)  \end{pmatrix} \hspace{1cm} \hbox{(\rm layers 1, $\cdots$ N)}
\end{equation}

\subsection{Inverse iteration algorithm}
In order to solve Eq.(\ref{eigenpb}) we use a generalization of the inverse
iteration method \citep[see][]{Dupret2001}. Starting from a first estimate of
the eigenvector $\rm Y_0$ and eigenfrequency correction $\delta \sigma_0$ (we take $\delta \sigma_0=0$, i.e.
$\sigma=\sigma_0$), we compute the next step in the iteration using the formula:
\begin{align}
\rm Y_{\rm k+1} \, = \, \rm AA^{-1} \, \rm AA_{\delta \sigma} \, \rm Y_k 
\end{align}
provided $\rm AA^{-1} \, \rm AA_{\delta \sigma}$ is diagonalizable. Solving the
eigenvalue problem is then equivalent to solving the linear
system:
\begin{align}
\rm AA \, \rm Y_{\rm k+1} \, = \, \rm AA_{\delta \sigma}\, \rm Y_k 
\end{align}
To do so, we perform a LU factorization of the matrix $\rm AA \, = \, \rm L \,
U$, where L is a lower triangular matrix and U an upper triangular one.
Therefore, the LU factorization needs to be made only once, and the only thing
to be done afterwards is to solve at each step of the inverse iteration the two
triangular systems:
\begin{align}
\rm L \, X \, &= \, \rm AA_{\delta \sigma} Y_k\\
\rm U \, Y_{k+1} \, &= \, \rm X \nonumber
\end{align}
To avoid ill-conditionning problems, we adopt a special kind of pivoting
strategy where the pivoting is done alternatively on the columns and on the
lines of the matrix.
 Thanks to the chosen finite difference resolution (Sect.
3.4), AA and AA$_{\delta \sigma}$ are block diagonal matrices, therefore, L and
U are also block diagonal matrices, where the blocks are triangular. The
non-zero elements are all located inside the blocks and, during the algorithm,
the permutations between lines and the permutations between columns will keep
the non-zero elements in the same block. This allows us to keep
the narrow band shape for the system, and reduces the memory needed and the
computational time. Once the eigenvector Y is computed with sufficiently high
precision, the eigenvalue can be calculated using a generalization of the
Rayleigh ratio:
\begin{align}
\delta \sigma \, = \, \frac{\rm Y^* \,  AA_{\delta \sigma}^* \, AA \, Y}{\rm Y^*\, AA_{\delta \sigma}^* \, AA_{\delta \sigma} \, Y}
\label{Rayleigh_ratio}
\end{align}
where $\rm Y^*$ and $\rm AA_{\delta \sigma}^*$ are the hermitian conjugate of Y
and $\rm AA_{\delta \sigma}$. The value of $\delta \sigma$ given by Eq.
(\ref{Rayleigh_ratio}) minimizes:
\begin{align}
S^2=\mid \mid (AA - \delta \sigma AA_{\delta \sigma} ) Y \mid \mid^2 
\end{align}
The solution is then obtained when a certain iteration criterium is reached,
which depends on the precision we want on the oscillation frequency.

\section{Conclusion and perspectives}
We presented here a new code that performs 2D non-perturbative calculations of
adiabatic oscillations for all kinds of stellar structures --
stratified, differentially rotating. We used a numerical method which
efficiently saves computationnal time and memory, and would make the code
available for seismic interpretation. It is now under a series of test
-- comparison with perturbative methods for evolved stellar models, and
comparison with non-perturbative computations for
polytropic models. With this new tools, we aim at modeling oscillations of 2D
stratified models of stars. A first step will be to take as the equilibrium model a 1D evolved stellar model where the whole effect of rotation on the microphysics have been taken into account, and use a self-consistent method \citep[used in][for example]{Roxburgh2006} in order to compute the distorsion due centrifugal acceleration.  

\acknowledgements
DRR is supported by the CNES (``Centre National d'Etudes Spatiales'') through a
postdoctoral fellowship.

\end{document}